\documentclass[prd,nofootinbib,showpacs,twocolumn]{revtex4-1}
\usepackage{color}
\usepackage{bm}
\usepackage{graphicx}
\usepackage{amsmath}
\usepackage{amssymb}
\usepackage{enumitem}
\usepackage{hyperref}
\usepackage{subfigure}
\usepackage{array}
\usepackage[english]{babel}
\usepackage{tensor}
\usepackage{comment}
\usepackage[normalem]{ulem}
\usepackage[dvipsnames]{xcolor}

\def\be{\begin{equation}}
\def\ee{\end{equation}}
\def\ba{\begin{eqnarray}}
\def\ea{\end{eqnarray}}
\def\l{\left}
\def\r{\right}
\def\f{\frac}

\allowdisplaybreaks

\begin{document}

\title{Can $f(Q)$-gravity challenge $\Lambda$CDM? }

\author{Lu\'is Atayde, Noemi Frusciante}
\affiliation{
\smallskip
 Instituto de Astrofis\'ica e Ci\^{e}ncias do Espa\c{c}o, Faculdade de Ci\^{e}ncias da Universidade de Lisboa, Edificio C8, Campo Grande, P-1749016, Lisboa, Portugal }

\begin{abstract}
We study observational constraints on the non-metricity $f(Q)$-gravity which reproduces an exact $\Lambda$CDM background expansion history while modifying the evolution of linear perturbations. To this purpose  we use Cosmic Microwave Background (CMB) radiation, baryonic acoustic oscillations (BAO), redshift-space distortions (RSD), supernovae type Ia (SNIa), galaxy clustering (GC) and weak gravitational lensing (WL) measurements. 
We set stringent constraints on the parameter of the model controlling the modifications to the gravitational interaction at linear perturbation level. 
We find the model to be statistically preferred by data over the $\Lambda$CDM according to the $\chi^2$ and deviance information criterion statistics for the combination with CMB, BAO, RSD and SNIa.  This is mostly associated to a better fit to the low-$\ell$ tail  of CMB temperature anisotropies.

\end{abstract}


\date{\today}

\maketitle

\section{Introduction} \label{sec:Intro}

The scientific goal of ongoing and next generation of cosmological surveys  is to understand the true nature of the cosmic acceleration which relies on testing the standard cosmological model, $\Lambda$ cold-dark-matter ($\Lambda$CDM), and any deviation from it. Still considering the cosmological constant $\Lambda$ as the main source of this phenomenon, one can construct gravity theories which are indistinguishable from $\Lambda$CDM at the background level but showing interesting and distinguishable signatures on the dynamics of perturbations.  In the following we will investigate whether there exists a gravity theory with these features able to challenge the  $\Lambda$CDM scenario.  

We will consider an extension of the  Symmetric Teleparallel General Relativity, the $f(Q)$-gravity, for which gravity is attributed to the non-metricity scalar $Q$~\cite{Nester:1998mp,Dialektopoulos:2019mtr,Jimenez:2019ovq,Lu:2019hra,Lu:2019hra,Bajardi:2020fxh,Jimenez:2019ovq}.  Detailed investigations of this theory have been performed in  many directions \cite{Lazkoz:2019sjl,Ayuso:2020dcu,Barros:2020bgg,Dialektopoulos:2019mtr,Jimenez:2019ovq,Bajardi:2020fxh,Flathmann:2020zyj,DAmbrosio:2020nev,Frusciante:2021sio,Khyllep:2021pcu,Anagnostopoulos:2021ydo}.  If we require the expansion history to match the one of 
$\Lambda$CDM, the functional form of $f(Q)$ is selected and can be derived analytically \cite{Jimenez:2019ovq}.  Then precise and measurable effects can be identified on  the matter density power spectrum, the Cosmic Microwave Background (CMB) radiation angular power spectrum and the lensing spectrum \cite{Frusciante:2021sio}. Cosmological constraints on this model are limited to background probes \cite{Ayuso:2020dcu} and when considering Redshift Space Distortion (RSD) data, constraints are obtained only on the additional parameter of the $f(Q)$-model and on the amplitude of the matter power spectrum at present time and scale of 8 h$^{-1}$Mpc, $\sigma_8^0$ (while fixing the base cosmological parameters to the $\Lambda$CDM best fit values)~\cite{Barros:2020bgg}. According to RSD data the so called $\sigma_8$ tension \cite{DiValentino:2020vvd} between Planck and Large Scale Structure data is alleviated for this $f(Q)$ model. 

In this work we provide for the first time cosmological constraints by means of Markov chain Monte Carlo (MCMC) methods and we use large sets of  data  spanning from measurements of the  background expansion of the Universe to those of gravitational potentials, matter density and temperature fluctuations power spectra.   We conclude our investigation with   a model selection analysis which will inform us whether the $f(Q)$ model analysed is supported by data over the $\Lambda$CDM scenario.

\section{The Model}\label{Sec:Model}

The action for the $f(Q)$-gravity can be written as follows~\cite{BeltranJimenez:2017tkd}
\be \label{eq:action}
S=\int d^4x\sqrt{-g}\l\{-\frac{1}{2\kappa^2}\l[Q+f(Q)\r]+\mathcal{L}_m(g_{\mu\nu},\chi_i)\r\}\,,
\ee
where $g$ is the determinant of the metric $g_{\mu\nu}$,  $\kappa^2 = 8 \pi G_N$ with $G_N$ being the Newtonian constant, 
$Q$ is the non-metricity scalar and it is defined as 
 $Q=-Q_{\alpha \mu \nu}P^{\alpha \mu \nu}$. The latter expression includes the non-metricity tensor $Q_{\alpha \mu \nu}$, which reads $Q_{\alpha \mu \nu}=\nabla_\alpha g_{\mu \nu}$ and $P^{\alpha}_{\phantom{\alpha}\mu\nu}=-L^{\alpha}_{\phantom{\alpha}\mu\nu}/2+\l(Q^\alpha-\tilde{Q}^\alpha\r)g_{\mu\nu}/4-\delta^\alpha_{(\mu}Q_{\nu)}/4$, where $Q_\alpha=g^{\mu\nu}Q_{\alpha\mu\nu}$, $\tilde{Q}_\alpha=g^{\mu\nu}Q_{\mu\alpha\nu}$ and $L^\alpha_{\phantom{\alpha}\mu\nu}=(Q^\alpha_{\phantom{\alpha}\mu\nu}-Q_{(\mu\nu)}^{\phantom{(\mu\nu)} \alpha})/2$. The action includes also a general function of the non-metricity scalar $f(Q)$ and the Lagrangian, $\mathcal{L}_m$, of standard matter fields, $\chi_i$. 

Let us note that in flat space the action \eqref{eq:action} has been shown to be equivalent to General Relativity (GR) for $f(Q)=0$~\cite{BeltranJimenez:2019tjy}. Thus in this context any deviation from GR can be cast in $f(Q)$.

We will now consider a background defined by the flat Friedmann-Lema{\^i}tre-Robertson-Walker (FLRW) line element:
\begin{equation}
    ds^2=-dt^2+a(t)^2\delta_{ij}dx^idx^j\,,
\end{equation} 
where $a(t)$ is the scale factor and $t$ is the cosmic time. It can be shown that on a FLRW background the non-metricity scalar becomes  $Q = 6H^2$ \cite{BeltranJimenez:2017tkd,Jimenez:2019ovq} where as usual we define $H\equiv\dot{a}/a$ as the Hubble parameter. Here the dot stands for a derivative with respect to $t$. 

The modified Friedmann equations can then be derived and have the form~\cite{BeltranJimenez:2017tkd}
\begin{eqnarray}
  &&    \label{eq:FriedEq}
    H^2 + 2 H^2 f_Q - \frac{1}{6} f = \frac{\kappa^2}{3} \rho_{i} ,  \\
  && (12H^2f_{QQ}+f_Q+1)\dot{H}=-\frac{\kappa^2}{2}(\rho_{m}+p_{ m})\,,
\end{eqnarray}
where  $f_{Q}\equiv\partial f/\partial Q$, $f_{QQ}\equiv\partial^2 f/\partial Q^2$ and $\rho_{m}$ and $p_{m}$  are respectively the energy density and pressure of the matter components. The latter satisfy the continuity equation for perfect fluids, $\dot{\rho}_m+3H(\rho_m+p_m)=0$. 

 In this work we  select the form of the $f(Q)$ function in such a way the main source of cosmic acceleration is  $\Lambda$, in doing so we can  investigate how modifications appearing only at the level of the perturbations can impact the cosmological constraints. We note that this is indeed a common practice \cite{Song:2006ej,Pogosian:2007sw,Zhao:2010qy,Hojjati:2015ojt,Bag:2018jle}. Then, assuming we want to mimic the $\Lambda$CDM background evolution, the form of the $f(Q)$ function can be analytically obtained from the first Friedmann equations and it is~\cite{Jimenez:2019ovq}:
\be\label{model}
f(Q) =  \alpha H_0\sqrt{Q} + 6H_0^2 \Omega_\Lambda,
\ee
where $\alpha$ is a dimensionless constant, $H_0$ is the present day value of the Hubble parameter and  $\Omega_\Lambda$ is the energy density parameter of the cosmological constant.  

The $\alpha$ parameter does not enter in the evolution of the expansion history by construction but it can largely affect the dynamics of the linear matter perturbations and gravitational potentials $\Phi(t,x_i)$ and $\Psi(t,x_i)$. 
This will allow us to investigate whether the inclusion of one additional parameter, defining the  deviation from GR, can lead to a better fit to data \footnote{Usually when more parameters  are included in the parameterization defying the deviation from GR, there are certain disadvantages:  it  might lead to lose the constraining power of data \cite{Salvatelli:2016mgy} and the increasing complexity (i.e. the number of free parameters) which in principle can lead  to  a better fit with the data needs to be  accurately evaluated using a model selection analysis  in order to provide the model assessment in comparison to the standard cosmological model.   The parametrization in Eq. (\ref{model}) considers one extra parameter and as such will reduce these contingencies.}.

Considering the Newtonian gauge, the perturbed line element,  around the FLRW background, is 
 \be
 ds^2=-(1+2\Psi)dt^2+a^2(1-2\Phi)\delta_{ij}dx^idx^j\,.
 \ee 
 Furthermore assuming the quasi-static approximation, it can be shown that for $f(Q)$-gravity the two gravitational potentials coincide, $\Phi=\Psi$, as in GR~\cite{Jimenez:2019ovq}. However the Poisson equation, which defines the relation between  the linear matter perturbations, $\delta \rho_m$, and the gravitational potentials, in Fourier space reads ~\cite{Jimenez:2019ovq}: 
 \be\label{eq:Poisson}
-k^2\Psi= 4 \pi \f{G_N}{1+f_Q}a^2\rho_m \delta_m\,,
\ee
where $\delta_m\equiv\delta\rho_m/\rho_m$ is the density contrast.
Therefore $f_Q$ modifies the strength of the gravitational interaction towards an effective gravitational coupling $\mu=1/1+f_Q$.
 
In a recent work \cite{Frusciante:2021sio} it has been shown that an effective gravitational coupling of the form  (\ref{eq:Poisson}) has measurable and interesting features on cosmological observables which strongly depend on the sign of $\alpha$. In detail, for $\alpha<0$, the gravitational interaction is stronger than in GR ($\mu>1$), then the growth factor is suppressed and the matter power spectrum is predicted to be enhanced compared to the one of the $\Lambda$CDM (sharing the same cosmological parameters). For the same reason it enhances the lensing power spectrum, being the lensing gravitational potential defined as $\phi_{lens}=(\Phi+\Psi)/2$. A  time variation of the latter impacts the late-time Integrated Sachs-Wolfe (ISW) effect whose signature is a suppressed low-$\ell$ tail of the temperature-temperature power spectrum. A completely opposite behavior is instead found when $\alpha>0$ which corresponds to a weaker gravity (see \cite{Frusciante:2021sio} for details).

\begin{table*}[th!]
\centering
\begin{tabular}{|l|l|l|l|l|l|}
\hline
 Model &   $\alpha$ & $n_s$ & $H_0 $ & $\Omega_m^0 $ & $\sigma_8^0 $ \\ \hline \hline
 $\Lambda$CDM (PLK18)& - &  $0.97\pm 0.01   $ & $68.0\pm 1.4        $ & $0.31\pm 0.02   $ & $0.85\pm 0.04   $  \\
 $\Lambda$CDM (PBRS)& - &  $0.970^{+0.008}_{-0.007}$& $68.1\pm 0.80     $ & $0.30\pm 0.01   $& $0.843^{+0.032}_{-0.037}   $\\
 $\Lambda$CDM (PBRSD)&  - &  $0.970\pm{0.008}$ & $68.33^{+0.76}_{-0.77}     $ & $0.302^{+0.010}_{-0.0097}  $ &  $0.829\pm0.031$   \\ \hline
 $f(Q)$ (PLK18)& $-0.64^{+0.64}_{-0.60}     $ &   $0.97\pm 0.01  $ & $68.3^{+1.5}_{-1.4}        $ & $0.304\pm 0.019   $ & $0.848^{+0.038}_{-0.037}   $  \\
 $f(Q)$ (PBRS)& $-0.56^{+0.58}_{-0.57}     $ &   $0.968^{+0.007}_{-0.008}$ & $68.14^{+0.79}_{-0.84}     $ & $0.305^{+0.011}_{-0.010}   $ &  $0.839^{+0.032}_{-0.031}   $  \\
 $f(Q)$ (PBRSD)& $-0.05^{+0.34}_{-0.36}$ & $0.970^{+0.008}_{-0.007}$ & $68.35\pm 0.80     $ & $0.302\pm 0.010   $ & $0.828\pm0.032   $  \\ \hline
\end{tabular}
\caption{Marginalised constraints on cosmological and model parameters at 95\% C.L. for the $\Lambda$CDM and $f(Q)$ models. }
\label{Tab:bounds}
\end{table*}

\begin{figure*}[th!]
\centering
\includegraphics[width=0.9\textwidth]{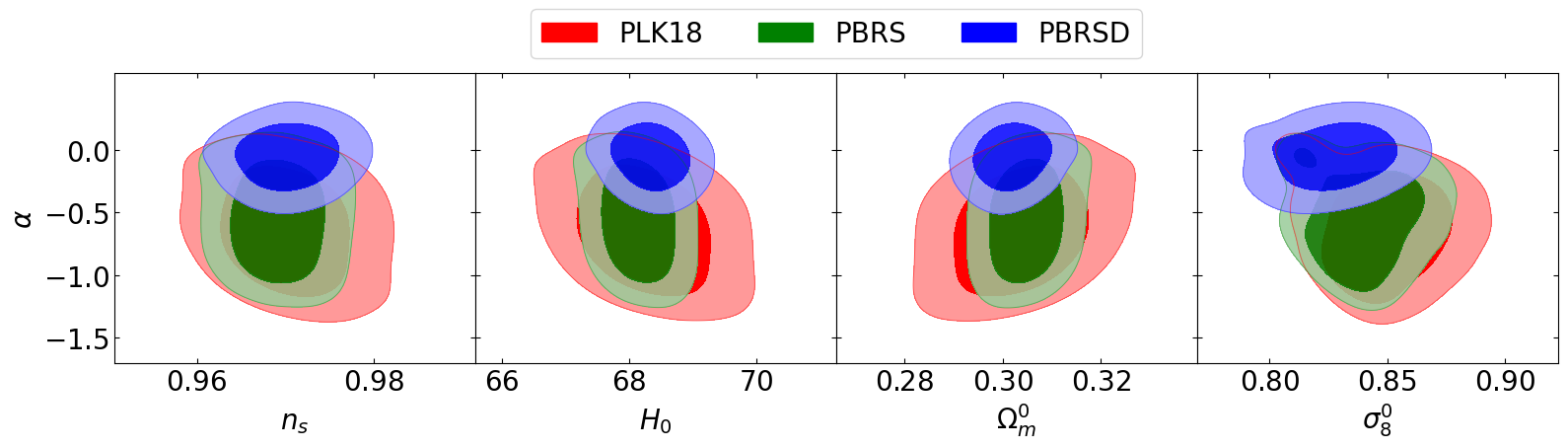}
\caption{Marginalised constraints at 68\% (darker) and 95\% (lighter) C.L. on the model parameter $\alpha$ and four cosmological parameters $H_0$, $n_s$, $\sigma_8^0$ and $\Omega_m^0$ obtained with the CMB data from Planck 2018 (PLK18, red), its combination with BAO, RSD and SNIa data (PBRS, green) and with DES data (PBRSD, blue).}\label{fig:constarints}
\label{figure1}
\end{figure*}


\section{Methodology and Data sets}\label{Sec:Data}
In the present cosmological analysis, we employ the Planck 2018 \cite{Planck:2019nip} (hereafter ''PLK18'') measurements of CMB temperature likelihood for large angular scales ($\ell =[2,29]$ for TT power spectrum) and for the small angular scales a joint of TT, TE and EE likelihoods ($\ell =[30,2508]$ for TT power spectrum, $\ell =[30,1996]$ for TE cross-correlation and EE power spectra). 

We then include baryonic acoustic oscillation (BAO) data from
the 6dF Galaxy Survey \cite{Beutler:2011hx} and from the Sloan Digital Sky Survey (SDSS) DR7 Main Galaxy Sample \cite{Ross:2014qpa}.
Furthermore, we  consider the combined BAO and RSD datasets from the SDSS DR12 consensus release \cite{BOSS:2016wmc}.

We complement the dataset with  the Joint Light-curve Array (JLA) of Supernova Type IA from the  Supernova Legacy Survey (SNLS) and SDSS \cite{SDSS:2014iwm}. We will consider the joint analysis with PLK18+BAO+RSD+SNIa and we will refer to it as ''PBRS''.

Finally we will include  galaxy clustering (GC) and weak gravitational lensing (WL) measurements from the Dark Energy Survey Year-One (DES-1Y) data \cite{DES:2017myr}. We use a standard cut of the nonlinear regime following \cite{DES:2018ufa,Zucca:2019xhg}, because  we do not have a prescription for nonlinear corrections. We refer to this dataset simply as ''DES'' and we use it in combination with the previous data, hereafter ''PBRSD''. 

We use  a modified version of the Einstein-Boltzmann code \texttt{MGCAMB} \cite{Hojjati:2011ix,Zucca:2019xhg} in which the  $f(Q)$-model in Eq. (\ref{model}) has been implemented \cite{Frusciante:2021sio}.
For the MCMC likelihood analysis  we use the \texttt{MGCosmoMC} code \cite{Hojjati:2011ix}. We impose a flat prior on $\alpha \in [-3,3]$ and we vary the base cosmological parameters: the physical densities of cold dark matter $\Omega_c h^2$ and baryons $\Omega_b h^2$ (with $h=H_0/100$), the reionization optical depth $\tau$, the primordial amplitude $\ln(10^{10} A_s)$, the angular size of the sound horizon at recombination $\theta_{MC}$ and spectral index $n_s$ of scalar perturbations. We include 
massive neutrinos with a fixed total mass of $\Sigma m_\nu= 0.06$ eV.

\section{Results}\label{Sec:Constraints}

In Table \ref{Tab:bounds} we show the constraints at 95\% C.L.  of  a selection of the cosmological parameters $H_0$, $n_s$,  $\sigma_8^0$, $\Omega_m^0$ and of the parameter $\alpha$ for the $f(Q)$ model. For reference we include the results for the $\Lambda$CDM model also. In Fig. \ref{fig:constarints} we show the marginalized constraints at 68\% and 95\% C.L. for the $f(Q)$-model.

The $f(Q)$-model has cosmological parameters which are consistent with the $\Lambda$CDM scenario.  The weaker constraints are obtained with PLK18 only, but the joint analysis, with BAO, RSD and SNIa and then with DES, strengthen the bounds. This is particularly evident for  $H_0$. Additionally the bounds on $\alpha$ are compatible among the datasets and negative mean values are preferred in all cases, with PLK18  selecting the smaller value.  The reason is because  negative values of $\alpha$ suppress the large-scale temperature anisotropies  accommodating better the CMB data. However, the larger negative values of $\alpha$ allow for higher values of $\sigma_8^0$ as expected from the phenomenology of the model. Then    when including RSD and DES data  $\alpha$ moves towards  higher values  (less negative $\alpha$) and hence smaller values for $\sigma_8^0$. As a side effect, the exclusion of its larger values leads to narrower bounds  compared to PLK18.  When DES data are considered we can also notice that  positive values of $\alpha$  are also allowed at both 68\% and 95\% C.L.. This is because a positive  $\alpha$ suppresses the matter power spectrum compared to $\Lambda$CDM allowing for a lower $\sigma_8^0$, which is known to be preferred by DES Y1 data.  
In a previous work~\cite{Barros:2020bgg} it has been found  only positive values for $\alpha$ ($\alpha=2.0331^{+3.8212}_{-1.9596}$) using RSD data. This result is expected because such measurements allow for a lower growth rate of matter density perturbations, thus preferring $\alpha>0$. The main difference with our result is in the inclusion of a larger combination of data sets  and in particular of the CMB data which, as previously discussed, select the negative branch of $\alpha$.

\begin{table}[t!]
\centering
\begin{tabular}{|l|l|l|l|l|l|l|l|}
 \hline
Data & $\Delta \chi_{\rm eff}^2$ &
$\Delta DIC$ \\ \hline
 \hline
 PLK18 & -3.3820 & -4.4739 \\
PBRS & -2.9040 & -3.1317 \\
PBRSD & -2.3040 &  4.8203 \\
\hline
\end{tabular}
\caption{Results for the $\Delta \chi_{\rm eff}^2$ and $\Delta DIC$ obtained as the difference between the $f(Q)$ and $\Lambda$CDM scenarios. }
\label{Tab:statistics}
\end{table}

We conclude our analysis by computing the  Deviance Information Criterion (DIC)~\cite{RSSB:RSSB12062}, which will allow us to  quantify the preference of the $f(Q)$ model with respect to $\Lambda$CDM. The DIC is defined as 
\be
\text{DIC}:= \chi_\text{eff}^2 + 2 p_\text{D},
\ee
where $\chi_\text{eff}^2$ is the value of the effective $\chi^2$ corresponding to the maximum likelihood and $p_\text{D} = \overline{\chi}_\text{eff}^2 - \chi_\text{eff}^2$, with the bar being the average of the posterior distribution. The DIC accounts for both the goodness of fit (through the $\chi_\text{eff}^2$) and for the bayesian complexity of the model (with $p_\text{D}$), disfavoring more complex models.  Models with smaller DIC should be preferred to models with larger DIC.  See Refs. \cite{Liddle:2009xe,Peirone:2019aua,Peirone:2019yjs,Frusciante:2019puu,Frusciante:2020gkx,Anagnostopoulos:2021ydo,Rezaei:2021qpq} for applications to alternative cosmological scenarios. Therefore we define the following quantity
\be
\Delta \text{DIC} = \text{DIC}_\text{f(Q)} - \text{DIC}_\text{$\Lambda$CDM}\,,
\ee
which will indicate a preference for the $f(Q)$ model over the $\Lambda$CDM scenario if $\Delta \text{DIC}<0$. We show in Tab.~\ref{Tab:statistics}  the values for both the $\Delta \chi_{\rm eff}^2$ and  $\Delta DIC$, for each of the data sets we used. 

We notice that all the combinations of data sets employed produce a lower $\chi^2_{\rm eff}$ for the $f(Q)$ model compared to the standard cosmological scenario.  Thus $f(Q)$-gravity can fit the data better than $\Lambda$CDM. The better agreement with data is due to the ability of the $f(Q)$ scenario to lower the ISW tail of the TT power spectrum, as previously discussed. This is further proved by the negative $\Delta DIC$ values we obtain for the PLK18 data and its combination with BAO, RSD and SNIa, which show a significant support in favor of the $f(Q)$ model. When considering the most complete data set we realize that the preference is instead for the $\Lambda$CDM scenario. This is due to the inclusion of the DES data which prefers the larger mean value for $\alpha$  in order to have  a lower $\sigma_8^0$, thus  degrading the better fit to the low-$\ell$ tail of the TT power spectrum. Therefore in this case a better $\chi^2_{\rm eff}$ is not sufficient to compensate the bayesian complexity of the model introduced by the additional parameter $\alpha$. Let us note that in this analysis we have performed a cut at linear scales for the GC and WL measurements of DES Y1. Thus in order to draw any conclusion a further analysis including these scales is necessary.

\section{Conclusion}\label{Sec:Conclusion}

We have provided stringent bounds at linear scales on  the cosmological and model parameters of the $f(Q)$-model defined in Eq. (\ref{model}). We have used a large sample of data including CMB, BAO, RSD, SNIa, WL and GC.  We have showed that for all combinations of data considered  the  $\chi_{\rm eff}^2$ statistics indicates that the $f(Q)$ model can fit better the data compared to the standard cosmological scenario, due to the ability of the model to lower the ISW tail compared to $\Lambda$CDM. The DIC statistical criterion significantly favors the $f(Q)$-model over $\Lambda$CDM when  PLK18 and PBRS are employed, while the DES data support the latter. In this case we stress that a further investigation is required in order to extend our analysis to nonlinear scales. We also note that the recent results from DES Y3 \cite{DES:2021wwk} show  a better agreement on $\sigma_8$ with Planck data compared to DES Y1. In this regards it can be informative to reconsider the $f(Q)$ model when the DES Y3 data will be available. 

Given the compelling features of the $f(Q)$ model, it can be counted among the challenging candidates \cite{Peirone:2019aua,Frusciante:2019puu,Anagnostopoulos:2021ydo} to the $\Lambda$CDM scenario.   It would be of interest to compare these scenarios among each other to find under the same conditions (datasets, priors, methodology, etc) which is the most promising model.  We leave this investigation for the future.

\begin{acknowledgments}
The authors thank M. Benetti, M. Martinelli and S. Peirone for useful discussions.
This work is  supported by Funda\c{c}\~{a}o para a  Ci\^{e}ncia e a Tecnologia (FCT) through the research grants UIDB/04434/2020, UIDP/04434/2020, PTDC/FIS-OUT/29048/2017, CERN/FIS-PAR/0037/2019 and  FCT project ``CosmoTests -- Cosmological tests of gravity theories beyond General Relativity" with ref.~number CEECIND/00017/2018.

\end{acknowledgments}

\appendix


\bibliographystyle{aipnum4-1}
\bibliography{biblio}

\end{document}